\documentclass[aps,prl,reprint]{revtex4-2}
\usepackage{graphicx}
\usepackage{textcomp}
\usepackage{amssymb}
\usepackage[utf8]{inputenc}
\usepackage{amsmath}
\usepackage{CJKutf8}
\usepackage{tipa}
\usepackage{comment}
\usepackage{mathtools}
\usepackage{braket}
\usepackage{listings}
\usepackage{xcolor}
\usepackage{soul}
\usepackage{etoolbox}
\usepackage{lineno}
\usepackage{float}
\usepackage{ulem}

\renewcommand{\vec}[1]{\mathbf{{#1}}}
\definecolor{Darkgreen}{rgb}{0,0.4,0}
\definecolor{listinggray}{gray}{0.9}
\definecolor{lbcolor}{rgb}{0.9,0.9,0.9}
\usepackage[colorlinks=true,linkcolor=blue,citecolor=blue,urlcolor=black]{hyperref}
\begin{document}

\title{Suppression of coherent light scattering in a three-dimensional atomic array
}

\author{{Yu-Kun Lu}}
\altaffiliation{These authors contributed equally to this work.}
\affiliation{Department of Physics, Massachusetts Institute of Technology, Cambridge, MA 02139, USA}
\affiliation{Research Laboratory of Electronics, Massachusetts Institute of Technology, Cambridge, MA 02139, USA}
\affiliation{MIT-Harvard Center for Ultracold Atoms, Cambridge, MA, USA}

\author{Hanzhen Lin
(\begin{CJK*}{UTF8}{gbsn}林翰桢\end{CJK*})}
\altaffiliation{These authors contributed equally to this work.}
\affiliation{Department of Physics, Massachusetts Institute of Technology, Cambridge, MA 02139, USA}
\affiliation{Research Laboratory of Electronics, Massachusetts Institute of Technology, Cambridge, MA 02139, USA}
\affiliation{MIT-Harvard Center for Ultracold Atoms, Cambridge, MA, USA}

\author{{Jiahao Lyu}}
\affiliation{Department of Physics, Massachusetts Institute of Technology, Cambridge, MA 02139, USA}
\affiliation{Research Laboratory of Electronics, Massachusetts Institute of Technology, Cambridge, MA 02139, USA}
\affiliation{MIT-Harvard Center for Ultracold Atoms, Cambridge, MA, USA}

\author{Yoo Kyung Lee}
\affiliation{Department of Physics, Massachusetts Institute of Technology, Cambridge, MA 02139, USA}
\affiliation{Research Laboratory of Electronics, Massachusetts Institute of Technology, Cambridge, MA 02139, USA}
\affiliation{MIT-Harvard Center for Ultracold Atoms, Cambridge, MA, USA}

\author{{Vitaly Fedoseev}}
\affiliation{Department of Physics, Massachusetts Institute of Technology, Cambridge, MA 02139, USA}
\affiliation{Research Laboratory of Electronics, Massachusetts Institute of Technology, Cambridge, MA 02139, USA}
\affiliation{MIT-Harvard Center for Ultracold Atoms, Cambridge, MA, USA}

\author{Wolfgang Ketterle}
\affiliation{Department of Physics, Massachusetts Institute of Technology, Cambridge, MA 02139, USA}
\affiliation{Research Laboratory of Electronics, Massachusetts Institute of Technology, Cambridge, MA 02139, USA}
\affiliation{MIT-Harvard Center for Ultracold Atoms, Cambridge, MA, USA}

\begin{abstract}   
    Understanding how atoms collectively interact with light is not only important for fundamental science, but also crucial for designing light-matter interfaces in quantum technologies. Over the past decades, numerous studies have focused on arranging atoms in ordered arrays and using constructive (destructive) interference to enhance (suppress) the coupling to electromagnetic fields, thereby tailoring collective light-matter interactions. These studies have mainly focused on one- and two-dimensional arrays. However, only three-dimensional (3D) arrays can demonstrate destructive interference of coherent light scattering in all directions. This omnidirectional suppression of coherent light scattering in 3D atomic arrays has thus far not been experimentally demonstrated. Here, we observe a strong reduction of light scattering in a 3D atomic array prepared in the form of a Mott insulator in optical lattices. The residual light scattering is shown to be caused by the delocalization of atoms, Raman scattering, and inelastic scattering associated with saturation. We also demonstrate how light scattering can be a sensitive probe for density fluctuations in many-body states in optical lattices, enabling us to characterize the superfluid-to-Mott insulator phase transition as well as defects generated by dynamical parameter ramps. The results of our work can be used to prepare subradiant states for photon storage and probe correlations for many-body systems in optical lattices.
\end{abstract}
\maketitle

When many atoms emit or scatter light, they all interact with the same modes of the electromagnetic field. Therefore, light scattering can be collective, which results in superradiant enhancement or subradiant suppression of the emitted light~\cite{PhysRev.93.99,PhysRevA.2.883,gross1982superradiance}. Of particular interest are periodic arrays of atoms where optical lattices and tweezers are used to control the geometry of individual atoms~\cite{gross2017quantum,kaufman2021quantum}. Recent work includes light scattering by linear chains (1D) into free space~\cite{glicenstein_collective_shift_chain_2020} or into a waveguide/cavity~\cite{PhysRevLett.117.133603, PhysRevLett.117.133604, yan_cavity_superradiant_2023}, light scattering off a two-dimensional (2D) Mott insulators~\cite{christof_microscope_scatter_2011,rui_sub_radiant_mirror_2020}, and Bragg scattering in three-dimensional (3D) optical lattices~\cite{PhysRevLett.75.2823,PhysRevLett.75.4583,PhysRevLett.107.175302,hart2015observation,shao2024antiferromagnetic}. There are many theoretical studies of collective emission in atomic arrays (see Refs.~\cite{mewton2007radiative,zoubi2010metastability,sutherland2016collective,PhysRevA.106.053717,PhysRevResearch.4.023207,masson2022universality,fayard_optical_2023,masson_dicke_multilevel_2024} and the references therein), and proposals on harnessing cooperative light scattering in 1D or 2D arrays for single photon nonlinear optics~\cite{PhysRevLett.116.103602, shahmoon_cooperative_resonance_2017} or for designing light-matter interfaces~\cite{chui2015subwavelength,facchinetti2016storing,asenjo2017exponential,manzoni2018optimization,needham2019subradiance,bekenstein2020quantum,masson2020atomic,patti2021controlling,PhysRevA.104.063707}. 

\begin{figure}
    \centering
    \includegraphics[width=\linewidth]{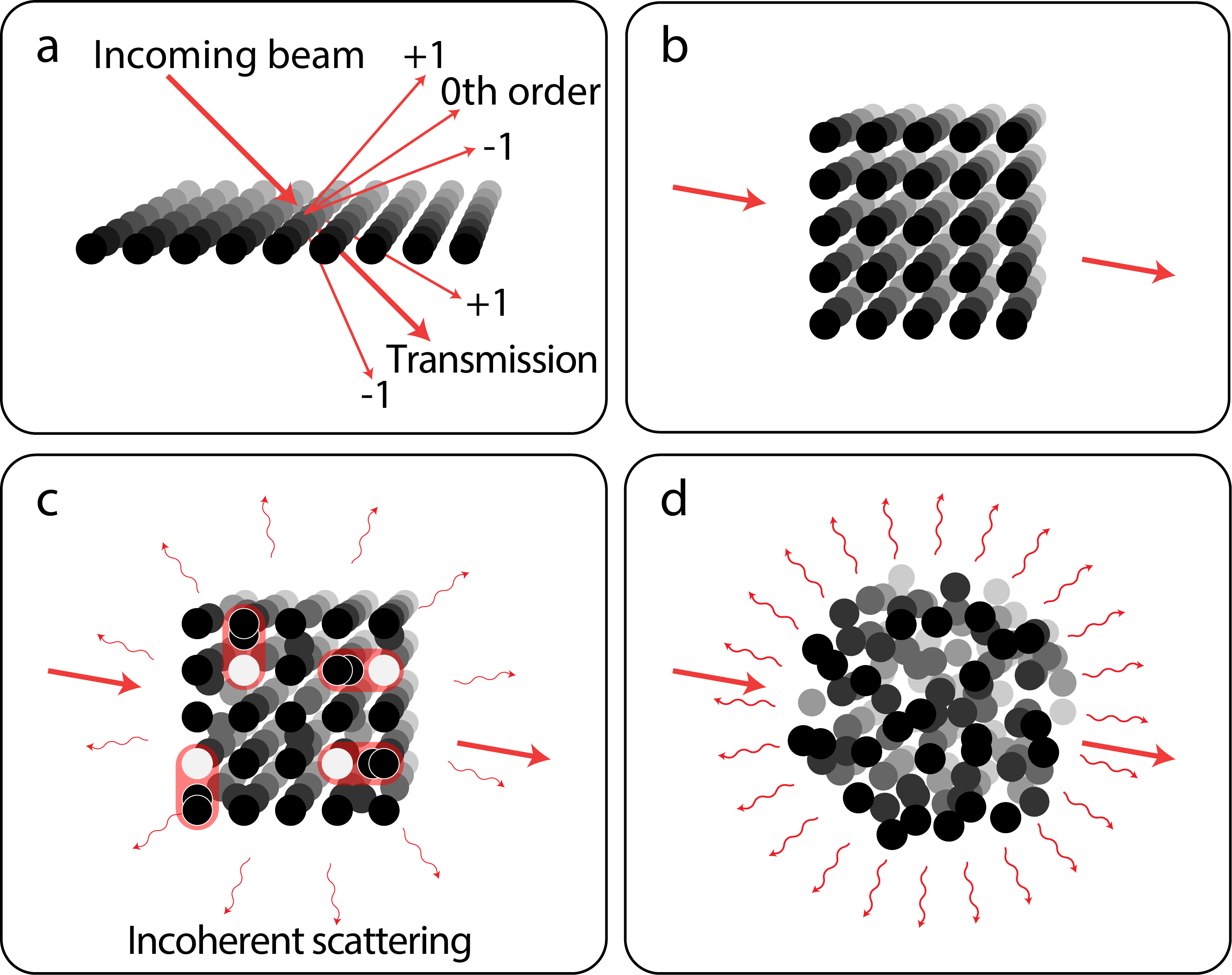}
    \caption{\textbf{Light scattering in 2D and 3D atomic arrays.} \textbf{a}, When the spacings between atoms are larger than $\lambda/2$, the coherent light scattering in 2D (and lower dimensions) has constructive interference in several diffraction orders and destructive interference elsewhere, maintaining the total scattering intensity. \textbf{b}, In 3D arrays, destructive interference happens at all angles as long as the probe beam is not close to a Bragg angle. Therefore, the total coherent light scattering intensity is suppressed. Defects (\textbf{c}) or disorder (\textbf{d}) reduce the suppression of light scattering.
    }
    \label{Fig1}
\end{figure}

\begin{figure*}
    \centering
    \includegraphics[width=2\columnwidth]{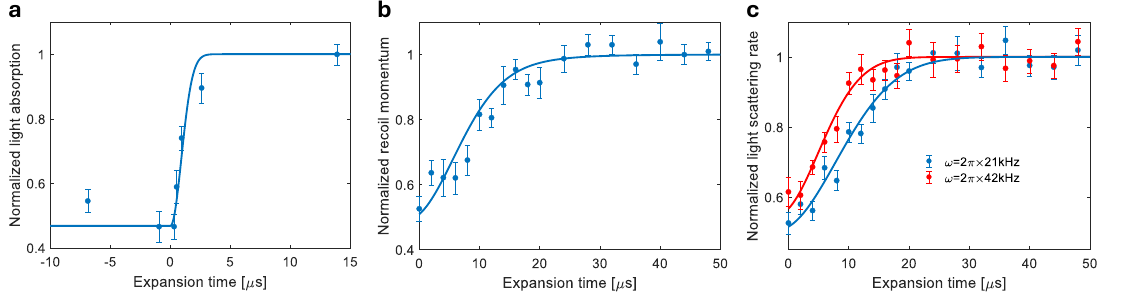}
    \caption{\textbf{Suppression of coherent scattering in 3D atomic array.} 
    \textbf{a}, Suppression of coherent light scattering measured by detuned absorption imaging of $^7$Li in 3D lattice. After switching off the lattice, the rms size of the atomic wavepacket expands as a function of time: $\Delta x(t)=\Delta x(0)\sqrt{1+\omega^2 t^2}$ leading to the normalized light scattering $I(t)=1-A \exp(-t^2/\tau^2)$. Here, $A$ is the suppression factor before expansion and $\tau=\sqrt{2m/\hbar \omega}/Q$ the time constant of the expansion from a lattice with trap frequency $\omega$~\cite{fedoseev2024coherent}. After long enough expansion, the suppression disappears as the scattering becomes fully incoherent. The solid curve is the theoretical prediction without any free parameters. 
    \textbf{b}, Suppression of photon scattering momentum recoil of $^{162}$Dy atoms released from a 3D lattice. The solid curve is the theory prediction with the suppression amplitude as a fitting parameter (see Methods). 
    \textbf{c}, Suppression of coherent light scattering measured by fluorescence imaging of $^{162}$Dy in 3D lattice. The blue (red) data points represent measurements for lattice trapping frequency $\omega=2\pi\times21\,\text{kHz}$ ($2\pi\times42\,\text{kHz}$). The solid lines represent the expected exponential shape with time constant $\tau$ obtained from independent measurements of the trap frequencies, and the suppression amplitude $A$ as a fitting parameter. The reduced suppression in deep lattices comes from the incoherent light scattering caused by the light-induced dipolar interaction on multiply occupied sites.
 Figures \textbf{a}, \textbf{b}, and \textbf{c} show the disappearance of light scattering suppression as the wavepackets expand in free space. The error bars throughout the whole paper are purely statistical and reflect the s.e.m.
      }
    \label{Fig2}
\end{figure*}

Previous studies have primarily investigated light scattering in 1D or 2D periodic atomic arrays. However, there is a fundamental difference for light scattering in a 3D atomic array compared to lower dimensions (Fig.~\ref{Fig1}). The coherent light scattering intensity $I(\vec{Q})$ with momentum transfer $\vec{Q} = \vec{k}_{\rm{in}}-\vec{k}_{\rm{out}}$ is proportional to $N$ times the structure factor $S(\vec{Q})=\left\lvert \sum_{j}^N\exp{(-i\vec{Q}\cdot\vec{R}_j)}\right\rvert^2/N$~\cite{footnote_coh}. Here $N$ is the number of atoms and $\vec{R}_j$ are the positions of the atoms. This implies momentum conservation: a photon can change its momentum by $\hbar\vec{Q}$ only when the density distribution or the density correlation function has a Fourier component at wavevector $\vec{Q}$. 
For two dimensions, there is always scattered light since the structure factor only enforces quasi-momentum conservation of the two in-plane components of the wavevector of the scattered light. The wavevector component perpendicular to the surface is not conserved and can ensure that the magnitude of the total wavevector of the scattered light fulfills the dispersion relation for light at the incident frequency. For example, any 2D array of atoms is phase-matched for reflection. However, in a 3D lattice, momentum transfer is possible only in units of reciprocal lattice vectors. This will not satisfy energy conservation unless the incident light fulfills the so-called Bragg condition. As a result, a perfect crystal in 3D will not scatter any light unless a specific Bragg condition is met.

The scattered light intensity $I(\vec{Q})$ fulfills a sum-rule: its integral over all $\vec{Q}$ is the sum of the single-atom contributions and is independent of the arrangement of the atoms, as long as the density distributions of the atoms are not overlapping (see Methods). 
For lower-dimensional arrays the integration over transverse $\vec{Q}$ is approximated by an integration over scattering angles. According to the sum rule, the total scattered intensity is independent of the arrangement of the atoms~\cite{footnote}. In contrast, the same sum rule in three dimensions requires integration over all three components of $\vec{Q}$. The total scattering rate for single-frequency incident light integrates only over a spherical surface in the three-dimensional $\vec{Q}$-space. Integration over all $\vec{Q}$-space can be physically realized only with a white broadband spectrum of the incident light, which involves all frequencies of the scattered light.

The suppression of coherent light scattering from densely filled 3D arrays allows highly sensitive studies of processes that cause incoherent scattering. Incoherent scattering probes density fluctuations and thus reveals crucial information about correlations and defects in many-body systems~\cite{Helmut_PRL,Helmut_PRA,PhysRevLett.103.170404,PhysRevA.81.013404,PhysRevA.92.013613,PhysRevA.81.013415,PhysRevA.84.033637}. Despite the conceptual simplicity and significant implications, neither the suppression of light scattering in 3D atomic arrays and its use to characterize density fluctuations in such systems has been experimentally demonstrated (except for our recent work on which-way information when expanding atomic wavepackets scatter light~\cite{fedoseev2024coherent}).

In this work, we use ultracold atoms to prepare a 3D array in the form of a Mott insulator, and observe a strong reduction of scattered light.
We show that the coherent scattering is strongly suppressed, and the residual incoherent scattering originates from both single-body and many-body effects. The single-body contribution is due to inelastic scattering that can come from the delocalization of the atoms, the existence of Raman scattering, or the saturation of the optical transition~\cite{footnote_coh}. The many-body contribution is caused by density fluctuations (defects) in the sample, which can originate from the nature of the many-body quantum state in equilibrium or the excitations created by dynamical parameter ramps. In particular, we show that by freezing the density fluctuations as defects in the array, we can use incoherent light scattering as a probe of the superfluid-to-Mott insulator phase transition due to the different nature of their density fluctuations~\cite{Helmut_PRL,Helmut_PRA}. We also studied the creation of defects by loading the lattice at different speeds, which can be described by the Kibble-Zurek mechanism~\cite{kibble1976topology,zurek1985cosmological,del2014universality}.
Our method is a general technique to probe density fluctuations of many-body systems in optical lattices.

We prepare our 3D Mott insulator by adiabatically loading $3\times10^4$ $^{162}$Dy (or $^{7}$Li) atoms from a Bose-Einstein condensate (BEC) into a 3D rectangular optical lattice. 
Light scattering is studied by exciting the atoms using a far-detuned probe beam and collecting the scattered photons at a 90-degree angle relative to the probe beam. The measurement is perturbative (around 0.15 photons per atom were scattered for $^{162}$Dy, 0.04 for $^7$Li), so the sample remains unchanged during the probe pulse. The detuning of the probe beam is chosen to be around 100 linewidths away from resonance for $^{162}$Dy (10 linewidths for $^7$Li) to minimize effects from multiple scattering and the shift/broadening of the resonance caused by light-induced electric dipolar interactions~\cite{guerin2017light}.

We measure the suppression of Rayleigh scattering by comparing the number of photons scattered in a 3D lattice with the number after a sufficiently long ballistic expansion, when each atom delocalizes and coherent scattering vanishes. This technique to distinguish coherent and incoherent light is the tool used throughout this paper. Fig.~\ref{Fig2}a demonstrates reduced absorption for the detuned probe light by around 50~\%. According to the optical theorem, the light absorption is equal to the light scattered over all solid angles. Our result therefore demonstrates the overall suppression of coherent light scattering integrated over all angles. At the same time, the suppression of absorption also implies a reduction of momentum recoil caused by light scattering, which can be measured from the shift of cloud center after a long time-of-flight (Fig.~\ref{Fig2}b). For all other measurements in this paper, we use fluorescence detection, which is almost background-free (Fig.~\ref{Fig2}c). Since we are far away from any Bragg condition, the same suppression is observed for absorption, momentum recoil, and fluorescence. 

In the Methods section, we calculate the structure factor for our geometry of finite size and estimate the role of an imperfect Mott insulator (which has defects in the form of holes). With a typical structure factor of 0.1 in our experiment, 90~\% of the elastically scattered light is suppressed, and therefore almost all of the observed light scattering of our 3D arrays is due to inelastic light scattering. 
\begin{figure}
    \centering
    \includegraphics[width=\columnwidth]{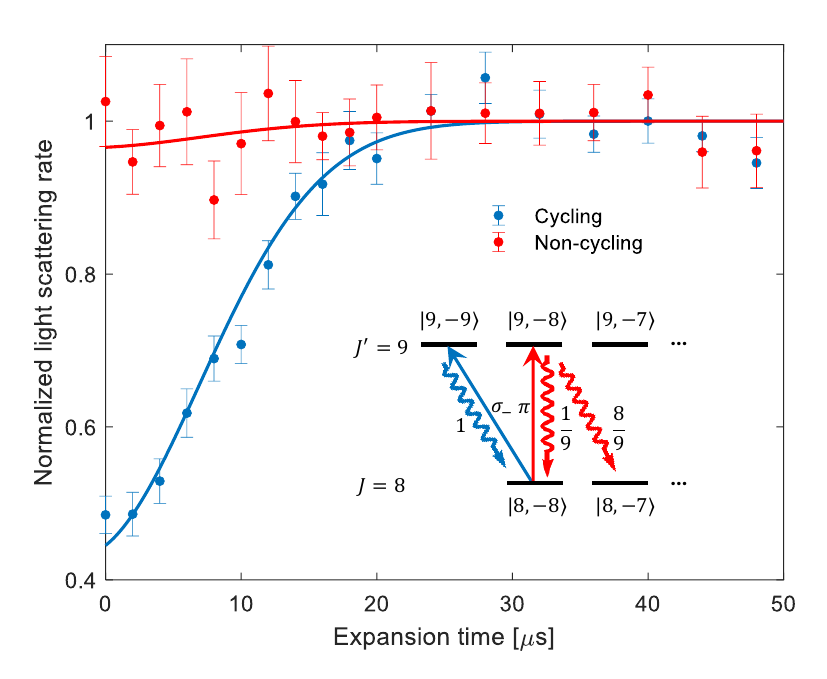}
    \caption{\textbf{Incoherent scattering caused by Raman scattering.} Here we compare light scattering for a cycling transition ($\ket{J=8,m_J=-8}\rightarrow \ket{J'=9,m_{J'}=-9}$) (blue) and non-cycling transition ($\ket{J=8,m_J=-8}\rightarrow \ket{J'=9,m_{J'}=-8}$) (red) using the 626~nm transition of $^{162}$Dy. The inset shows the polarization and branching ratios for the two transitions. The solid curves are fits to the data, from which the suppression factors $A$ are obtained. 
    The branching ratio of the $\pi$ transition is 8:1, but since the coherently scattered light is not detected at our detection angle (due to the dipolar emission pattern), we expect to observe only incoherent light, consistent with a suppression factor of $A=0.03(4)$. The cycling case has a suppression factor of $A=0.57(3)$. }
    \label{Fig3}
\end{figure}

For a single two-level atom and weak probe laser intensity, the elastic scattering fraction is given by the Debye-Waller factor $D=\exp(-Q^2\Delta x^2)$, with $\Delta x$ being the rms size of the atomic wavepacket along $\vec{Q}$. The inelastic fraction $1-D$ which is around 50~\% for our lattice depth is due to motional excitation of the atom from the photon recoil. 
In previous work, we studied how the Debye-Waller factor decreased during time-of-flight expansion~\cite{fedoseev2024coherent} and discussed that the deeper reason for inelastic light is the (partial) entanglement between scattered photons and the state of the atom. The theory curves in Fig.~\ref{Fig2} show that the elastic light scattering during time-of-flight is in quantitative agreement with the ballistic expansion of Gaussian wavepackets, initially trapped at each lattice site. The sensitivity to the size of the atomic wavefunction can be used for thermometry. This is demonstrated in the Methods section, where we used a pulse of light to heat up the atoms to higher bands, and then monitored the resulting energy or temperature increase by the decrease in elastic light scattering. In this case, we are using band excitations for thermometry. As we discuss below, particle and hole excitations also increase light scattering, and can be used for thermometry~\cite{PhysRevLett.103.170404,PhysRevA.84.033637}.

Elastic light scattering requires the atom to return to the initial state; otherwise, the scattered photon is entangled with the final state of the atom, and the phase of the photon alone is random. In addition to the excitation of motional states, we now demonstrate this effect by spin excitations. For this, we excite the atom on a non-cycling optical transition and show that the Raman scattered light is inelastic (see Fig.~\ref{Fig3}).

\begin{figure}
    \centering
    \includegraphics[width=\columnwidth]{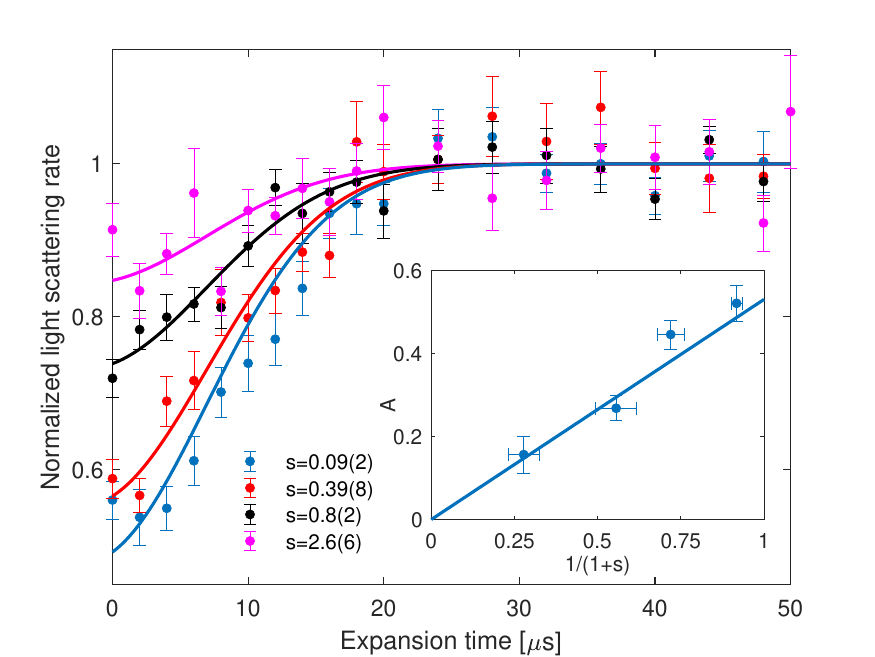}
    \caption{\textbf{Incoherent scattering caused by saturation.} Light scattering of $^{162}$Dy is measured for different saturation parameters $s$ (different colors). The number of scattered photons is kept constant for the datasets with different $s$ by adjusting the pulse durations and detunings accordingly. The solid curves are fits to the data, and the suppression factor $A$ is obtained from the fits. The suppression factor $A$ obeys a linear dependence on the coherent scattering fraction $1/(1+s)$, as shown in the inset.}
    \label{Fig4}
\end{figure}

\begin{figure*}
    \centering
    \includegraphics[width=2\columnwidth]{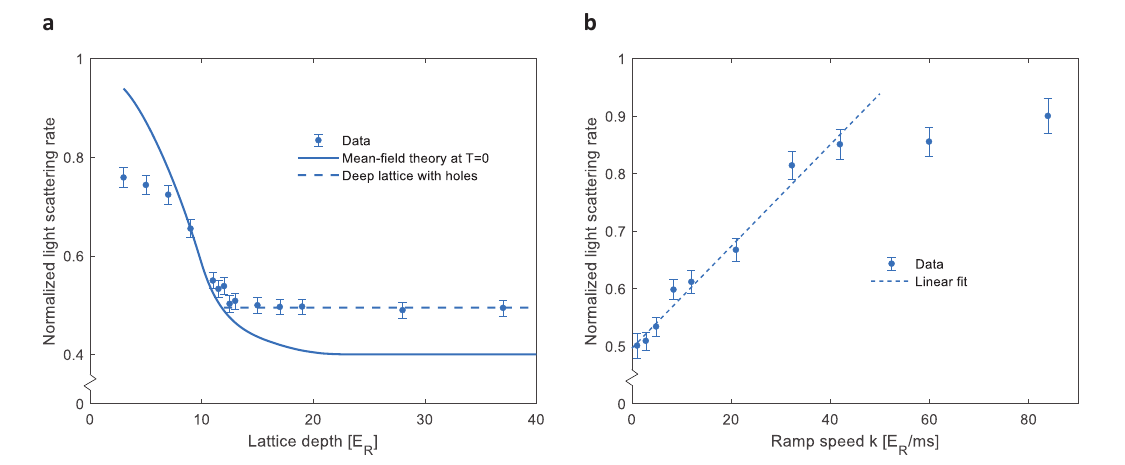}
    \caption{\textbf{Incoherent scattering as a probe of density fluctuations.} 
    \textbf{a}, \ Light scattering of $^{7}$Li atoms in 3D optical lattice across the superfluid-to-Mott insulator phase transition. The light scattering decreases as the lattice depth increases, and eventually plateaus when the system enters the Mott insulator phase. The solid line is the Gutzwiller (mean-field) calculation, with the coherent scattering fraction $D=0.6$. The dashed line contains contributions from scattering off holes and other effects~\cite{fedoseev2024coherent}. \textbf{b}, Light scattering measured for different ramp speeds from superfluid-to-Mott insulator phase. The light scattering initially grows linearly, consistent with the Kibble-Zurek mechanism and eventually saturates for fast enough ramps because the density fluctuations are frozen in.
    }
    \label{Fig5}
\end{figure*}

Until now, we have considered the weak excitation limit. When the optical transition is saturated, multi-photon scattering events emerge, leading to the Mollow triplet. The sidebands correspond to transitions between the different dressed states and are therefore also inelastic, whereas the carrier is mostly elastic (unless strongly saturated). According to Ref.~\cite{API}, the inelastic fraction of the scattered light is $s/(1+s)$ where $s=2(\Omega/\Gamma)^2/(1+4\Delta^2/\Gamma^2)$ is the saturation parameter, $\Omega$ the Rabi frequency of the probe beam, $\Delta$ the detuning, and $\Gamma$ the linewidth of the transition. By varying the saturation parameter of the probe beam, we observed that the suppression of light scattering before expansion (defined as the suppression factor $A$) shows a linear dependence on $1/(1+s)$, which is the predicted elastic fraction of the scattered light (Fig.~\ref{Fig4})~\cite{PhysRevLett.116.183002}. 

Including all three possible mechanisms discussed above, we arrive at the elastic scattering fraction $f$ for single atoms $f=\eta D/(1+s)$ with $\eta$ 
being the branching ratio in the observation direction. For a many-body system with an arbitrary filling pattern in the lattice, the normalized light scattering is related to the structure factor $I=f S(\vec{Q})+(1-f)=1-A$~\cite{fedoseev2024coherent}, where $A=f(1-S(\vec{Q}))$ is the suppression factor introduced before. The small structure factor of regular arrays implies that light scattering is a sensitive diagnostic for density fluctuations. We will now describe two experiments where the density fluctuations are caused either by the nature of the many-body quantum state or by defects generated during a dynamical parameter ramp.

In our system, atoms in the optical lattice are governed by the Bose-Hubbard model, which hosts the superfluid phase and Mott insulator phase at low temperatures~\cite{sachdev1999quantum}. The two key parameters in the Bose-Hubbard model are the tunneling rate $t$ and the on-site interaction strength $U$. When the $U$ is weak compared to $t$, atoms can tunnel across different lattice sites, establish phase coherence, and create density fluctuations, forming the superfluid phase. When $U$ is strong compared to $t$, the tunneling will be energetically suppressed, as will the phase coherence and density fluctuations; thus, the system enters the Mott insulator phase. There is a second-order quantum phase transition between the superfluid phase and the Mott insulator phase at $U/t\approx29.3$~\cite{PhysRevB.75.134302}. 

The $^7$Li atoms in a 3D optical lattice were prepared near the many-body ground state at different lattice depths. Then, the lattice depth was quenched to the maximum value (37 $E_{\rm{R}}$) to freeze in the density fluctuations as defects in the array. In the case of a homogeneous sample with negligible density-fluctuation correlations between different lattice sites, the structure factor is simply $S(\vec{Q})\approx\braket{\delta\hat{n}^2}/\braket{\hat{n}}$ where $\hat{n}$ represents the occupation of each site (see Methods). When the lattice depth is low, the system is in the superfluid phase, and each site is approximately a coherent state with a Poisson distribution of occupation number. As a result, the light scattering approaches the value of a random sample ($S(\vec{Q})\rightarrow1$). When the lattice depth is high, the system enters the Mott insulator phase with a Fock state on each site, and the density fluctuation approaches 0 ($S(\vec{Q})\rightarrow0$). We compare our observations (Fig.~\ref{Fig5}a) to the predictions of the zero-temperature Gutzwiller calculation (see Methods) and obtain qualitative agreement. For high lattice depth, the data deviated from the model by 10~\%, which can be fully accounted for by the presence of holes, surfaces, optical pumping, and saturation~\cite{fedoseev2024coherent}. 
For low lattice depths, the data deviates from the ideal theory by 20~\%. 
We predict that this is partially due to multiple occupancies on the same lattice site --- at low lattice depths, two-thirds of the atoms are on sites with a population of more than one atom where effects due to light-induced dipolar interactions become important~\cite{photo_association_review}. For instance, dipolar interactions can increase the fraction of inelastically scattered light and cause leakage out of the cycling transition by coupling the upper state to other Zeeman states.
However, while our simple models predict a reduction of light scattering, they cannot fully account for the observed 20~\% deviation.

Light scattering can also be used to obtain information about dynamical properties in many-body systems.
We study how a ramp through the quantum critical point of the superfluid-to-Mott insulator transition can cause excitations.
The number of excitations generated by the ramp has a universal scaling with the ramp speed of the parameter~\cite{PhysRevB.72.161201,PhysRevB.86.144527}. We prepare the $^7$Li atoms in the superfluid phase at 5$E_{\rm{R}}$ lattice depth, ramp across the superfluid-to-Mott insulator phase transition at different speeds, and end at 37$E_{\rm{R}}$ to measure light scattering. If the ramp is slow, the system adiabatically follows the ground state of the Hamiltonian and becomes a nearly uniformly filled array with occupation $ n =1$ on each site. In this case, the structure factor is close to zero, i.e., there is no incoherent scattering from density fluctuations. As the ramp speed increases, excitations are created during the ramp and manifest as particle/hole defects in the array, which leads to an increase in incoherent scattering. We varied the ramp speed from 1$E_{\rm{R}}$/ms to 140$E_{\rm{R}}$/ms while the tunneling rate is $\approx2\pi \times0.5\,\rm{kHz}$ near $\approx10 E_{\rm{R}}$ when the transition is crossed, and observed that the incoherent scattering initially increased with the ramp speed and plateaued for ramp speed larger than 40$E_{\rm{R}}$/ms, indicating a timescale at which the local density fluctuations are frozen (Fig.~\ref{Fig5}b). The incoherent scattering is given by $S(\vec{Q})\approx\braket{\delta\hat{n}^2}/\braket{\hat{n}}\equiv n_{\rm{ex}}$ which directly reflects the defect density in the array. We observed that the defect density $n_{\rm{ex}}$ is consistent with the universal scaling with respect to the ramp speed given by the Kibble-Zurek mechanism of the on-tip phase transition in the superfluid-Mott insulator phase diagram with $n_{\rm{ex}}\propto k^{\alpha}$ and $\alpha=d\nu/(1+\nu z)=1$~\cite{sachdev1999quantum,PhysRevB.72.161201,PhysRevB.86.144527} (see Fig.~\ref{Fig5}b).

In conclusion, we have demonstrated the suppression of coherent scattering in 3D atomic arrays in all directions, which is fundamentally different from that of lower-dimensional arrays. We showed that the remaining incoherent scattering arises due to inelastic scattering from single atoms as well as density fluctuations in the many-body system. 3D atomic arrays can be used to investigate other collective light-matter interaction effects, such as superradiance and subradiance on atomic arrays~\cite{mewton2007radiative,zoubi2010metastability,sutherland2016collective,PhysRevA.106.053717,PhysRevResearch.4.023207,masson2022universality,fayard_optical_2023,masson_dicke_multilevel_2024,facchinetti2016storing,asenjo2017exponential,needham2019subradiance}. For example, the suppression of coherent light scattering in 3D atomic arrays can be used to prepare subradiant states if the $k$ vector of the spin-wave excitation generated by the probe beam can be shifted away from the phase-matched condition. This can be done either by an effective magnetic field gradient generated with a spin-dependent light shift~\cite{PhysRevLett.117.243601}, or by momentum transfer from two pi-pulses~\cite{PhysRevLett.115.243602,PhysRevLett.125.213602}. Our versatile method can investigate density fluctuations or perform thermometry in a broad range of many-body systems~\cite{Helmut_PRL,Helmut_PRA,PhysRevLett.103.170404,PhysRevA.81.013404,PhysRevA.92.013613,PhysRevA.81.013415,PhysRevA.84.033637}.

Our method of diagnosing many-body systems via light scattering complements existing optical microscopy methods, because spatially resolved imaging of atoms and coherent suppression of the total light scattering intensity cannot be achieved simultaneously. Therefore, our method will be especially useful when conventional optical microscopy cannot be applied, for example, in 3D samples or in systems with very short lattice spacing. Additionally, the perturbative nature of the light scattering probe makes our method less destructive (in particular, it does not suffer from the parity projection present in microscopy methods~\cite{QGMReview}), and can reveal additional information about the system. Furthermore, the degree of coherence is a qualitatively new observable in optical spectroscopy which should have broad applicability to both fundamental and applied studies.

\textit{Acknowledgments}.---We are grateful to Immanuel Bloch and Pascal Weckesser for discussions. 
We thank William R. Milner for proofreading the manuscript. We acknowledge support from the NSF (grant No. PHY-2208004), from the Center for Ultracold Atoms (an NSF Physics Frontiers Center, grant No. PHY-2317134), from the Vannevar-Bush Faculty Fellowship (grant no. N00014-23-1-2873), from the Gordon and Betty Moore Foundation GBMF ID \# 12405), from the Army Research Office (contract No. W911NF2410218) and from the Defense Advanced Research Projects Agency (award HR0011-23-2-0038). Yoo Kyung Lee acknowledges the MathWorks Science Fellowship. Yu-Kun Lu is supported by the NTT Research Fellowship.

%


\begin{thebibliography}{60}
\bibitem [{foo({\natexlab{a}})}]{footnote_coh}%
  {\bibinfo {title} {{Throughout this work, we define elastic (inelastic) light scattering as photon scattering processes that do not change (change) the internal or external quantum state of the atom. In contrast, coherent (incoherent) light scattering refers to processes where the scattered photons have fixed (random) phase relations with respect to the incoming photon. Inelastic scattering is always incoherent. In the absence of density fluctuation, elastic light scattering implies coherence. However, the light scattering caused by static density fluctuations is fully elastic but incoherent.}}} %
\bibitem [{foo({\natexlab{b}})}]{footnote}%
  {\bibinfo {title} {{The existence of the sum rule for 2D or lower can also be understood by the following thought experiment. Consider a microscope with a numerical aperture of 1 (therefore a spatial resolution of $\lambda/2$), with the atoms at its focus. If the spacing between the atoms is larger than $\lambda/2$, then their point-spread functions in the imaging plane have little overlap, so the interference of light emitted by different atoms is negligible. Therefore, the total light intensity emitted by the atoms remains the same regardless of the arrangement of the atoms. In 3D, the light scattered by atoms from different layers overlap in the image. This leads to the possibility of destructive interference and complete suppression of total light scattering.}}}%
\end{thebibliography}

\newpage
\clearpage

\setcounter{figure}{0}
\setcounter{equation}{0}
\makeatletter 
\renewcommand{\thefigure}{S\@arabic\c@figure}
\renewcommand{\theequation}{S\@arabic\c@equation}

\makeatother

\onecolumngrid

\section*{Methods}
\subsection{Experimental procedures and modeling}

\textbf{Dysprosium} The $^{162}$Dy atoms are confined in a rectangular optical lattice formed by three retro-reflected laser beams. Light scattering is performed using the 626~nm optical transition of $^{162}$Dy with linewidth $\Gamma=2\pi\times136\,\rm{kHz}$. The probe beam propagates in the horizontal plane at an angle of approximately $50^\circ$ with respect to one of the horizontal lattice beams. Scattered photons are collected along the vertical direction by an imaging system with a numerical aperture (NA) of 0.5. More information on the experimental configuration can be found in our previous work~\cite{fedoseev2024coherent}.

For fluorescence measurements throughout the paper (Fig.~\ref{Fig2}c, Fig.~\ref{Fig3}, Fig.~\ref{Fig4}), we apply two light pulses in every experimental run. The first pulse measures light scattering under different conditions, while the second pulse is done after a long (300~$\mu$s) expansion and serves as an atom count. The normalized light scattering rate is obtained by taking the ratio of the photon number collected in the two shots from the same run. 
We used $-12.7$~MHz detuning and 4~$\mu$s pulse duration for light scattering, except in Fig.~\ref{Fig4} where we changed both to maintain the same number of scattered photons for different saturation parameters $s$. 
The trapping frequencies for the atoms in all directions are 26~kHz for data in Fig.~\ref{Fig3} and Fig.~\ref{Fig4}. 
The model for light scattering time trace is $I(t)=1-A\exp(-t^2/\tau^2)$~\cite{fedoseev2024coherent}, and the suppression amplitude $A$ is the only fitting parameter. Because of the finite pulse duration $\Delta t$, all theory curves that assumed negligible pulse duration are shifted by $\Delta t/2$, since we are effectively measuring the light scattering at the midpoint of the light pulse.

For the momentum recoil measurement in Fig.~\ref{Fig2}b, the lattice constants are 532~nm in the horizontal directions and 266~nm in the vertical direction. The trapping frequencies are 24~kHz in the horizontal direction and 39~kHz in the vertical direction. For light scattering, we turn on light pulses with $-3.3$~MHz detuning for 4~$\mu$s immediately after releasing the atoms from the optical lattices. The recoil momentum is obtained by measuring the shift of the cloud center after a 6~ms time-of-flight. 
One key difference here compared to the fluorescence measurements is that the recoil momentum depends on light scattering into all directions, rather than the light scattering into a particular observation angle. The recoil is $\vec{p}(t)=\int d\Omega \vec{Q}f_{\rm{dip}}I(\vec{Q},t)$, with $\vec{Q}=\vec{k}_{\rm{in}}-\vec{k}_{\rm{out}}$ being the momentum transfer, $f_{\rm{dip}}$ being the dipole radiation pattern, and $I(\vec{Q},t)=S(\vec{Q})D(\vec{Q},t)+(1-D(\vec{Q},t))$ being the normalized light scattering for momentum transfer $\vec{Q}$. Using spherical coordinates with the $z$-axis along the quantization axis, we have $\vec{Q}=k\left(1-\sin\theta\cos\phi,\sin\theta\sin\phi,\cos\theta\right)$, $f_{\rm{dip}}=(1+\cos^2\theta)/2$, and $I(\vec{Q},t)=1-\exp(-\Delta x(t)^2Q_x^2-\Delta y(t)^2Q_y^2-\Delta z(t)^2Q_z^2)(1-S(\vec{Q}))$. We obtain the theory curve in Fig.~\ref{Fig2} by calculating the recoil momentum at different expansion times and using the suppression before expansion as the only fitting parameter.

\textbf{Lithium} The Mott insulator of $^7$Li atoms is prepared using a Feshbach resonance. After evaporating the atoms in a crossed dipole trap with scattering length $127 a_0$ (where $a_0$ is the Bohr radius), the atoms are transferred to an $1E_{\rm{R}}$ shallow optical lattice while the scattering length is tuned to $353 a_0$. This adjusts the density to match that of an $n=1$ Mott insulator, and sets the superfluid-to-Mott insulator transition at $\approx10E_{\rm{R}}$. Afterwards, unless stated otherwise, the lattice is ramped up over 40~ms to its final depth of $37E_{\rm{R}}$ using an exponential ramp. In the quench experiments, we ramp the lattice exponentially to an intermediate lattice depth in 40~ms, hold for 1~ms, then quench to $37E_{\rm{R}}$ in 0.1~ms. For the sweep experiments, we exponentially ramp the lattice to $5E_{\rm{R}}$ in 40~ms, then ramp the lattice linearly to $20E_{\rm{R}}$ with variable rate, then have a 1~ms hold before finally ramping to $37E_{\rm{R}}$ in 1~ms. After preparing the sample, we probe it using a 100~ns light pulse, which is 60~MHz detuned, similar to our previous work~\cite{fedoseev2024coherent}. For the absorption measurement in Fig.~\ref{Fig1}a is compared to a theoretical curve, which is obtained by averaging the Debye-Waller factor over the entire solid angle weighted by the dipolar emission pattern using $I_{\rm{abs}}(t)=\int d\Omega f_{\mathrm{dip}}I(Q,t)$.

\subsection{Light scattering as thermometry of the band excitations}
\begin{figure}
    \includegraphics[scale=0.58]{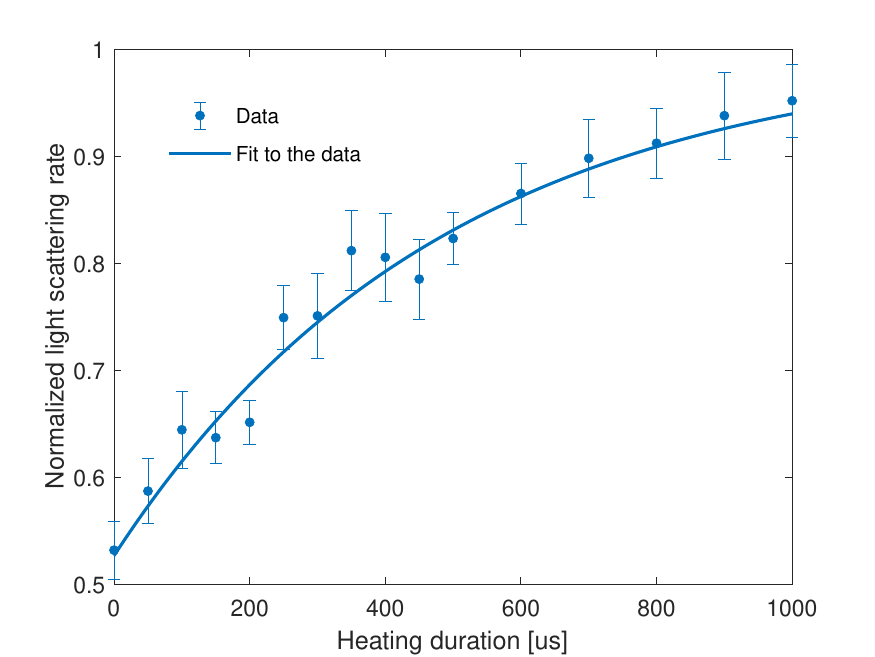}
    \caption{\textbf{Light scattering as thermometry for band excitations.} The heating pulse excites atoms to higher bands, which reduces the Debye-Waller factor and increases the total light scattering. The solid curve is a fit to the data $I(t)=1-A\exp(-t/t_0)$, with $t_0$ being the time constant determined by the heating rate. The fit gives $A=0.47(3)$ and $t_0=0.49(6)$ms.}
    \label{fig:FigS1}
\end{figure}

To demonstrate how light scattering can be used for thermometry, a heating light pulse with tunable duration is applied before the light scattering. The detuning is $2\Gamma_{\rm{421}}$ with respect to the 421~nm optical transition of dysprosium. The normalized light scattering rate is measured as a function of the heating duration (Fig.~\ref{fig:FigS1}). Because the photon recoil heating can transfer atoms to higher bands, the Debye-Waller factor is reduced, which leads to an increase in incoherent scattering (as shown in Fig.~\ref{fig:FigS1}).

\subsection{Sum rule of the structure factor}
The coherent light scattering with momentum transfer $\vec{Q}$ is proportional to the structure factor $S(\vec{Q})$:
\begin{equation}
\begin{aligned}
     S(\vec{Q}) & =\frac{1}{N}\left|\int \rho(\vec{r}) e^{-i \vec{Q} \cdot \vec{r}} d^n \vec{r}\right|^2 \\
& =\frac{1}{N} \int d^n \vec{r} d^n \vec{r}^{\prime} \rho(\vec{r}) \rho(\vec{r^{\prime}}) e^{-i \vec{Q}\left(\vec{r}-\vec{r}^{\prime}\right)}.
\end{aligned}
\end{equation}

Here $\rho(\vec{r}) =\sum_i \rho_i\left(\vec{r}-\vec{R}_i\right)$ is the density distribution of the $N$ atoms, and is given by the sum of the density distributions $\rho_i(\vec{r}-\vec{R}_i)$ of the individual atoms at position $\vec{R}_i$.

To prove the sum rule, we integrate $S(\vec{Q})$ over all $\vec{Q}$:
\begin{equation}
\label{sum_rule}
\begin{aligned}
\int S(\vec{Q}) d^n \vec{Q} & =\frac{1}{N} \int d^n \vec{r} d^n \vec{r}^{\prime}(2 \pi)^n \delta^{(n)} (\vec{r}-\vec{r}^{\prime}) \rho(\vec{r}) \rho\left(r^{\prime}\right) \\
& =\frac{(2 \pi)^n}{N} \int d^n \vec{r} \rho{(\vec{r})}^2.
\end{aligned}
\end{equation}

The square of the density can be expressed as:
\begin{equation}
\label{density}
    \rho(\vec{r})^2 =\sum_i \rho_i^2\left(\vec{r}-\vec{R}_i\right)+\sum_{i \neq j} \rho_i\left(\vec{r}-\vec{R}_j\right) \rho_j\left(\vec{r}-\vec{R}_i\right).
\end{equation}
For non-overlapping atoms, the second term vanishes. Therefore, the integral of coherent light scattering intensity $I_{\rm{coh}}(\vec{Q})=NS(
\vec{Q}
)$ over all $\vec{Q}$ is
\begin{equation}
    \int I_{\rm{coh}}(\vec{Q}) d^n \vec{Q} =N\frac{(2 \pi)^n}{N} \sum_i \int d^n \vec{r} \rho_i^2(\vec{r}) =\sum_i\int \left|\int \rho_i(\vec{r}) e^{-i \vec{Q} \cdot \vec{r}} d^n \vec{r}\right|^2d^n\vec{Q},
\end{equation}
which is the sum of the single-atom contributions and is independent of the arrangement of the atoms.

The derivation above is exact and holds for any dimension $n$, but it involves the integration over all frequencies of light. For atoms in 1D or 2D geometries embedded in 3D space, there is also a sum rule for total light scattering at a single frequency. In this case, the total light scattering into all solid angle $\int I(\vec{Q})d\Omega$ can be approximated by the integral over the in-plane $\vec{Q}$ within a finite region of size $\Delta Q\approx \lvert \vec{k_\mathrm{in}}\rvert$. Therefore, the $\delta$ function in Eq.~\ref{sum_rule} is replaced by a function with spatial extension $1/\Delta Q\approx \lambda$. As long as the atoms' density distributions are separated further than $\lambda$, the second term in Eq.~\ref{density} can be dropped, and we arrive at the sum rule for single-frequency light scattering. 
A similar argument applies if the integration over in-plane $\vec{Q}$ is cut off by the aperture of a lens. This proves the intuitive picture of a thought experiment, where a high-NA objective collects all scattered light to image the atoms, as we elaborated in the footnote~\cite{footnote}.

\subsection{Description of light scattering of atomic arrays}

\subsubsection{Debye-Waller factor} 
Photon scattering can be separated into two components based on whether the scattering changes the state of the atoms: when the atomic state is changed, light scattering is inelastic and does not interfere; if the atom returns to the original state, the scattered photons can interfere~\cite{fedoseev2024coherent,footnote_coh}.

Therefore, the total scattering intensity is given by
\begin{align}
    I=f S(\vec{Q})+(1-f).
\end{align}
Here $f$ describes the amount of inelastic scattering, and the structure factor $S(\vec{Q})$ captures the interference of the elastically scattered light, including static density fluctuations. $S(\vec{Q})$ is simply derived from the normalized square of the Fourier transform of the density distribution of all scatterers:

\begin{align}
    S(\vec{Q})=\lvert E(\vec{Q})\rvert^2/N;E(\vec{Q}) = \sum_j^N \exp(-i\vec{Q}\cdot\vec{R}_j).
\end{align}

\subsubsection{Structure factor for different geometries}

In the following, we consider light scattering far from Bragg angles from point-like scatterers arranged in a cubic lattice inside a specific volume.
We will show that for the cuboid geometry, light scattering is approximately equal to the light scattering from a single scatterer, while for a sphere it grows with the number of scatterers as $N^{2/3}$.\\

\noindent\textbf{1D arrays}.
For a 1D array, we can sum over all amplitudes $\vec{E}_j$ from $N$ scatterers in an array with spacing $a$:

\begin{align}
    I_\mathrm{1D} = \left\lvert \sum_j^N \vec{E}_j\right \rvert^2 =
    I_0\left\lvert \sum_j^N e^{-i\vec{Q}\vec{r}_j}\right \rvert^2     
    =I_0\left[\frac{\sin(Q_xNa/2)}{\sin(Q_xa/2)}\right]^2,    \label{eq:Sq_1D}
\end{align}
where $I_0=\lvert\vec{E}_j\rvert^2$ is the scattered intensity from an individual scatterer, $\vec{Q}$ is the momentum change of the scattered photons, and $Q_x$ is the projection of $\vec{Q}$ along the array axis . 

Averaging $I_{\rm{1D}}$ over a small interval of size $\delta Q_x\approx 1/(Na)$ around $Q_x$ in the direction far from the Bragg angles (where $\sin(Q_xa/2)\neq0$) will average out the numerator but not the denominator. We get  
\begin{align}
    \braket{I_\mathrm{1D}} \approx\frac{I_0}{2}\left[\frac{1}{\sin(Q_xa/2)}\right]^2,    
\end{align}
which is independent of $N$, on average equal to (up to a prefactor) $I_0$, the intensity of a single scatterer. Intuitively, only the edge of a 1D chain scatters light.\\

\begin{figure}
    \centering
    \includegraphics[width=\linewidth]{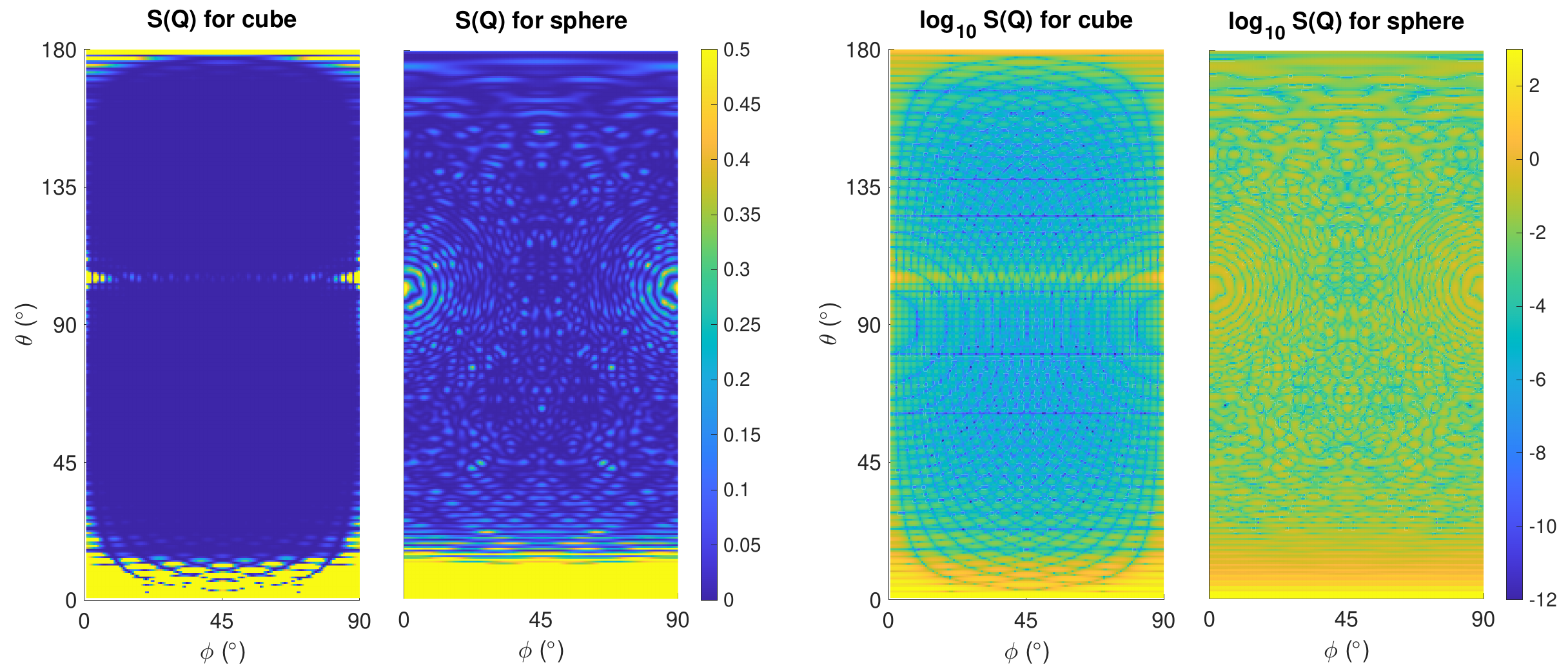}
    \caption{
    \textbf{Structure factors of a cubic lattice with cubic or spherical boundaries.}
    The incoming wavevector $\vec{k}_{\rm in}$ is fixed to $(0,0,k)$ for $k=2\pi/\lambda$ and $\vec{k}_{\rm out}$ is defined to be $k (\sin\theta\cos\phi,\sin\theta\sin\phi,\cos\theta)$. 
    The forward scattering angle $\theta=0$ always adds constructively with a maximum $S(\vec{Q})=N$ regardless of geometry.
    The left panel shows the structure factors for a cube and a sphere with $N\approx3\times 10^4$ atoms.
    The images are saturated to show small features.
    The right panel shows the structure factors of a cube and a sphere on a logarithmic scale.
    We note that there are Bragg angles at $(\theta\approx105^\circ,\phi=0,90^\circ)$.
    }
    \label{fig:FigS2}
\end{figure}

\noindent\textbf{3D cubic arrays}.
The structure factor $S(\vec{Q})$ for a cubic lattice is the product of the structure factors of three one-dimensional lattices:
\begin{align}
    S(\vec{Q})_{\rm cube}
    =\frac{1}{N} 
    \times \left[\frac{\sin(Q_xN_xa/2)}{\sin(Q_xa/2)}\right]^2 \left[\frac{\sin(Q_yN_ya/2)}{\sin(Q_ya/2)}\right]^2 \left[\frac{\sin(Q_zN_za/2)}{\sin(Q_za/2)}\right]^2
\end{align}
where $N=N_xN_yN_z$ is the total number of sites, $N_x$ ($N_y,N_z$) are the number of sites in the $x$ ($y,z$) directions, and $\vec{Q}=(Q_x,Q_y,Q_z)$ describes the momentum transfer along the axes of the cube. For a finite-size cube, the scattered intensity far away from Bragg angles is again $\propto I_0$, the intensity of a single scatterer, corresponding to $S(\vec{Q})_{\rm cube} = 1/N$.
An example for a cube of 32 sites on each edge is shown in Fig.~\ref{fig:FigS2}.

This result has an intuitive explanation: A rectangular 2D array of scatterers can be considered as a 1D array of effective scatterers, reducing to a new effective single scatterer. 
By analogy, a cuboid 3D array of $N$ scatterers produces light scattering equivalent to a single scatterer independent of $N$.\\

\noindent\textbf{Spherical samples}.
The experiment realizes approximately a spherical cloud of atoms. For a sphere, the resulting $S(\vec{Q})_{\rm sphere}$ differs slightly from that of the cube (Fig.~\ref{fig:FigS2}). 
In particular, for a finite sphere, the scattering intensity scales with the surface area of the sphere, as we derive now.

The scattering amplitude of the lattice with spherical boundaries $R<R_0$ is the Fourier transform of the real lattice 
\begin{align}
    E(\vec{Q}) = \mathcal{F}[{\rm sphere}\times\sum_j^N\delta({\vec{R}-\vec{R}_j})] = \mathcal{F}[{\rm sphere}]*\sum_j^N \exp(-i\vec{Q}{\cdot}\vec{R}_j)
\end{align}
where the multiplication becomes a convolution in Fourier space.
The scaling with radius $R_0$ comes only from the spherical factor.

The Fourier transform of a spherically symmetric function is also spherically symmetric:
\begin{align}
    \mathcal{F}[{\rm sphere}]=\frac{4\pi}{Q^3}[\sin(QR_0)-QR_0\cos(QR_0)]
\end{align}
In our case, $QR_0\gg1$, so we can ignore the smaller $\sin(QR_0)$ term. The total (coherent) scattering intensity due to surface effects from all scatterers is then
\begin{align}
    I_{\rm tot} = |E|^2 \propto (QR_0)^2\cos^2(QR_0) \propto N^{2/3}
\end{align}
where in the last step we have averaged the oscillating cosine term.

Intuitively, a 3D sphere consists of 1D chains with a range of different lengths; each such chain is an effective scatterer with a random phase, scattering light incoherently. 
The sphere has $\pi(R/a)^2$ such effective scatterers. Adding intensities from each 1D array results in
\begin{align}
I_{\mathrm{sphere}} = I_\mathrm{1D}\pi (R/a)^2 = I_\mathrm{1D}\pi(N/(4\pi/3))^{2/3}.
\end{align}
Therefore, a sphere containing $N$ scatterers scatters light with an intensity $\sim N^{2/3}$.

\begin{figure}[b]
    \includegraphics[scale=0.58]{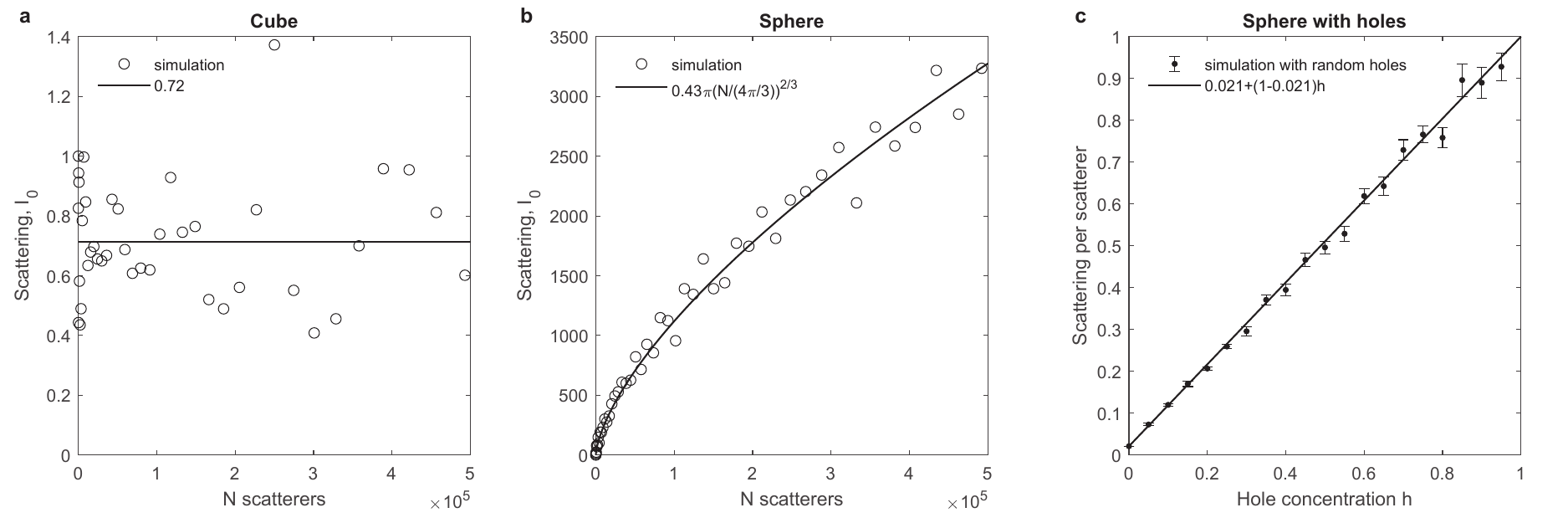}
    \caption{\textbf{Simulation of light scattering far from the Bragg angles from scatterers arranged on a cubic lattice inside a given geometrical shape.} Light scattering from a cube is independent of the number of scatterers (\textbf{a}). For a sphere, scattering is proportional to the surface area (\textbf{b}). In \textbf{a} and \textbf{b}, the data points are scattered due to the finite size of the solid angle for detection. (\textbf{c}) Simulation of light scattering by random holes. Error bars represent the random sampling of positions.}
    \label{fig:FigS3}
\end{figure}

\subsection{Defects in the lattice}
The discussions above assumed an array without defects. In our experiment, defects arise due to the density fluctuations of the many-body quantum states, defect generation in dynamical parameter ramps, or finite temperature effects. Here we describe the extra light scattering caused by the defects.

The electric field operator of the scattered light can be expressed by :
\begin{equation}
    \hat{E}\propto \sum_j \hat{n}_j e^{i\vec{Q}\cdot \vec{R}_j}.
\end{equation}
Therefore, the intensity operator is (the proportional factor is set to 1 for convenience):
\begin{equation}
    \hat{I}=\left|\hat{E}\right|^2=\sum_{j,k} \hat{n}_j\hat{n}_k e^{i\vec{Q}\cdot (\vec{R}_j-\vec{R}_k)}.
\end{equation}

The total light scattering is given by the expectation value of the intensity operator: 
\begin{equation}
I_\text{tot}=\braket{\hat{I}}=\braket{\left|\hat{E}\right|^2}=\sum_{j,k} \braket{\hat{n}_j\hat{n}_k} e^{i\vec{Q}\cdot (\vec{R}_j-\vec{R}_k)},
\end{equation}
while the coherent scattering is:
\begin{equation}
\label{coh}
    I_\text{coh}=\left|\braket{\hat{E}}\right|^2= 
    \left|\sum_j \braket{\hat{n}_j} e^{i\vec{Q}\cdot \vec{R}_j}\right|^2.
\end{equation}

To proceed, we further separate the density fluctuation from the mean density $\hat{n}_j=\braket{\hat{n}_j}+\delta \hat{n}_j$. Therefore, the total light scattering can be expressed as:

\begin{equation}
I_\text{tot}=\sum_{j,k} \braket{\hat{n}_j}\braket{\hat{n}_k} e^{i\vec{Q}\cdot (\vec{R}_j-\vec{R}_k)}+\sum_{j,k} \braket{\delta\hat{n}_j\delta\hat{n}_k} e^{i\vec{Q}\cdot (\vec{R}_j-\vec{R}_k)}=I_\text{coh}+\sum_{j} \braket{\delta\hat{n}_j^2}+\sum_{j\neq k} \braket{\delta\hat{n}_j\delta\hat{n}_k} e^{i\vec{Q}\cdot (\vec{R}_j-\vec{R}_k)}
\label{eq:density_fluctuation_scattering}
\end{equation}

The first term is the same expression as for coherent light scattering from a perfect 3-dimensional lattice, which vanishes away from the Bragg condition. The second term represents the light scattering caused by the on-site density fluctuation. The last term is the contribution from correlated density fluctuation on different sites, which can be ignored away from the superfluid-to-Mott insulator transition point. Therefore, the normalized light scattering (structure factor) in a homogeneous sample has the following simple expression $S(\vec{Q})\approx\braket{\delta\hat{n}^2}/\braket{\hat{n}}$.

Note that the coherent scattering only depends on the mean density distribution $\braket{\hat{n}_j}$, while the incoherent scattering comes from the density fluctuations $\braket{\delta\hat{n}_j\delta\hat{n}_k}$. The structure factor $S(\vec{Q})$ describes the total elastic scattering, which includes both the coherent and incoherent scattering described above.

We now apply this result to the situation of a regular crystal with holes. We assume that each site has a probability $h$ of being vacant and a probability $1-h$ of being occupied by one atom. The coherent scattering is given by $I_{\rm{coh}}=|E_0(0\times h+1\times (1-h))|^2=|E_0|^2(1-h)^2$ while the total scattering is $I_{\rm{tot}}=|E_0|^2(0\times h+1\times (1-h))=|E_0|^2(1-h)$. Here, $E_0$ represents the electric field of light scattered by a single atom. Therefore, the coherent scattering fraction is $1-h$. For a three-dimensional array with infinite size, the structure factor away from the Bragg condition is simply $S(\vec{Q})=h$ with $h$ being the hole concentration. The incoherently scattered light is $I_{\rm{tot}}-I_{\rm{coh}} = |E_0|^2 h (1-h)$ and shows the symmetry between light scattered by particles and by holes.

These conclusions about light scattering of a cube, a sphere, and a sphere with holes are consistent with simulations (Fig.~\ref{fig:FigS3}), where the light scattering is calculated by directly adding up the amplitudes from all sites for a given geometry. These simulations use the experimental parameters for the lithium setup with a lattice spacing of 532~nm and a wavelength of 671~nm for the probe light. The probe beam is sent in the direction [0 0 1], and light scattering is observed in the direction [1 1 0].

\subsection{Light scattering of bosons in optical lattices}

As derived in Eq.~\ref{eq:density_fluctuation_scattering}, light scattering of atoms in optical lattices can be affected by density fluctuations. Interacting bosonic atoms in optical lattices are described by the Bose-Hubbard Hamiltonian~\cite{sachdev1999quantum}:

\begin{equation}
\hat{H} = -t \sum_{\langle i,j \rangle} (\hat{a}_i^\dagger \hat{a}_j + \hat{a}_j^\dagger \hat{a}_i) + \frac{U}{2} \sum_i \hat{n}_i (\hat{n}_i - 1) - \sum_i (\mu-V_i) \hat{n}_i,
\end{equation}
where $t$ is the tunneling rate between nearest-neighbor sites, $U$ is the on-site interaction strength, $\mu$ is the chemical potential, and $V_i$ is the potential energy from the harmonic confinement of the optical lattice. Operators $\hat{a}_i^\dagger$ and $\hat{a}_i$ are the bosonic creation and annihilation operators at site $i$, and $\hat{n}_i = \hat{a}_i^\dagger \hat{a}_i$ is the occupation number operator at site $i$.

For bosons in optical lattices, the on-site number fluctuations can be calculated using the Gutzwiller (mean-field) approximation~\cite{sachdev1999quantum}. In the mean-field approximation, the tunneling term is decoupled by introducing the superfluid order parameter $\psi_i = \langle \hat{a}_i \rangle$:

\begin{equation}
\label{Decouple}
\hat{a}_i^\dagger \hat{a}_j \approx \langle \hat{a}_i^\dagger \rangle \hat{a}_j + \hat{a}_i^\dagger \langle \hat{a}_j \rangle - \langle \hat{a}_i^\dagger \rangle \langle \hat{a}_j \rangle = \psi_i^* \hat{a}_j + \hat{a}_i^\dagger \psi_j - \psi_i^*\psi_j.
\end{equation}

Substituting Eq.~\ref{Decouple} into the Hamiltonian, the mean-field Hamiltonian becomes:

\begin{align}
    \hat{H}_{\rm{MF}} & = \sum_{\langle i,j \rangle} \left[ -t  (\psi_i^* \hat{a}_j + \hat{a}_i^\dagger \psi_j - \psi_i^*\psi_j + {\rm{h.c.}})\right] + \sum_i \frac{U}{2} \hat{n}_i (\hat{n}_i - 1) - \sum_i (\mu-V_i) \hat{n}_i \\
    & = \sum_{i} \left[ -z t  (\hat{a}_i^\dagger \bar{\psi}_i+ {\rm{h.c.}}) + \frac{U}{2} \hat{n}_i (\hat{n}_i - 1) - (\mu-V_i) \hat{n}_i+zt\frac{(\psi_i^* \bar{\psi_i}+\rm{h.c.})}{2}\right]\\
    & = \sum_{i} \hat{H}_{\rm{i}}
\end{align}
Here $\hat{H}_{\rm{i}}$ is the single-site Hamiltonian, with $z$ being the coordination number and $\bar{\psi_i}=\frac{1}{z}\sum_{\braket{i,j}}\psi_j$ being the averaged order parameter around site $i$.
We can focus on single sites since the mean-field Hamiltonian is simply a sum of single-site Hamiltonians on all sites. The order parameter $\psi$ is determined self-consistently on each site. The mean-field ground state and the order parameter $\psi$ can be calculated by iterating this procedure. The on-site density fluctuations are obtained from the mean-field ground state. The total light scattering can then be obtained from the structure factor $S(\vec{Q})\approx\braket{\delta\hat{n}^2}/\braket{\hat{n}}$. Note that here we neglect the contribution of the correlations of density fluctuations on different sites to $S(\vec{Q})$, which should be a good approximation away from the critical point and for large momentum transfer $\vec{Q} \gg 1/\xi$ ($\xi$ is the healing length).

\begin{figure}
    \includegraphics[scale=0.58]{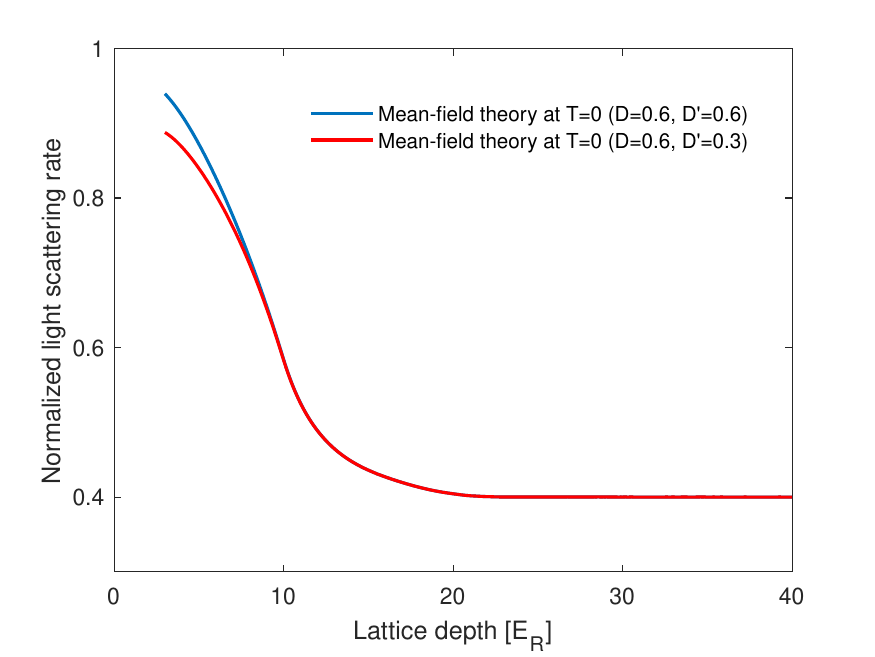}
    \caption{\textbf{Effect of incoherent scattering of higher occupation sites.} The normalized light scattering calculated from the zero-temperature mean-field theory with (without) the additional incoherent scattering is shown as the red (blue) curve. The Debye-Waller factor for singlons is fixed as $D=0.6$, while the coherent scattering fraction for $n>1$ sites is $D'=0.3$ (red curve) and $D'=0.6$ (blue curve).}
    \label{fig:FigS4}
\end{figure}

When there is more than one atom per site, the electric dipolar interaction on the excited state manifold leads to the possibility of exciting the relative motion, and can cause extra incoherent scattering~\cite{photo_association_review}. This can be modeled by introducing different Debye-Waller factors $D$ and $D'$ for occupations $n=1$ and $n>1$, respectively, with $D'<D$. 
The coherent scattering intensity of a single site is proportional to $|\braket{E}|^2\propto|\sum_n \sqrt{D'}np_n+(\sqrt{D}-\sqrt{D'})p_1|^2=|\sqrt{D'}\braket{\hat{n}}+(\sqrt{D}-\sqrt{D'})p_1|^2$, with $p_n$ being the probability to have $n$ atoms on the site. The total scattering is proportional to $\braket{|E|^2}\propto\sum_n p_n (D'n^2+(1-D')n)=D'\braket{\hat{n}^2}+(1-D')\braket{\hat{n}}$. Therefore, the normalized incoherent scattering is $(\braket{|E|^2}-|\braket{E}|^2)/\braket{\hat{n}}$. Fig.~\ref{fig:FigS4} compares the zero-temperature mean-field prediction with and without the reduction of coherence for higher occupation ($n>1$) sites. The additional incoherent scattering for $n>1$ sites causes a reduction of the total light scattering.

\end{document}